\documentclass[aip,jcp,amsmath,amssymb,preprint]{revtex4-1}

\usepackage{graphicx}
\usepackage{dcolumn}
\usepackage{bm}
\usepackage[version=3]{mhchem}

\begin{document}


\title{Quantum process tomography of entangled photons as a probe of intermediates of singlet fission in a tetracene derivative}



\author{Ajay~Ram~Srimath~Kandada}
    \thanks{These authors contributed equally to this work.}
    \affiliation{Center for Nano Science and Technology@PoliMi, Istituto Italiano di Tecnologia, via Giovanni Pascoli 70/3, 20133 Milano, Italy}

\author{Ilaria-Bargigia}
    \thanks{These authors contributed equally to this work.}
    \affiliation{School of Chemistry and Biochemistry, Georgia Institute of Technology, 901 Atlantic Drive NW, Atlanta, Georgia 30332, United~States}

 
 \author{Eric.~R.~Bittner}
    \affiliation{Department of Chemistry, University of Houston, Houston, Texas 77204, United~States}

\author{Carlos~Silva}
    \email[E-mail address: ]{carlos.silva@gatech.edu}
    \affiliation{School of Chemistry and Biochemistry, Georgia Institute of Technology, 901 Atlantic Drive NW, Atlanta, Georgia 30332, United~States}
    \affiliation{School of Physics, Georgia Institute of Technology, 837 State Street NW, Atlanta, Georgia 30332, United~States}
    \affiliation{School of Materials Science and Engineering, Georgia Institute of Technology, North Avenue, Atlanta, Georgia 30332, United~States }

\date{\today}

\begin{abstract}
Spin-entaglement has been proposed and extensively used in the case of correlated triplet pairs which are intermediate states in singlet fission process in select organic semiconductors. Here, we employ quantum process tomography of polarization entangled photon-pairs resonant with the excited state absorption of these states to investigate the nature of the inherent quantum correlations and to explore for an unambiguous proof for the existence of exciton entanglement.  
\end{abstract}


\maketitle

\begin{figure}
    \centering
    \includegraphics[width=7cm]{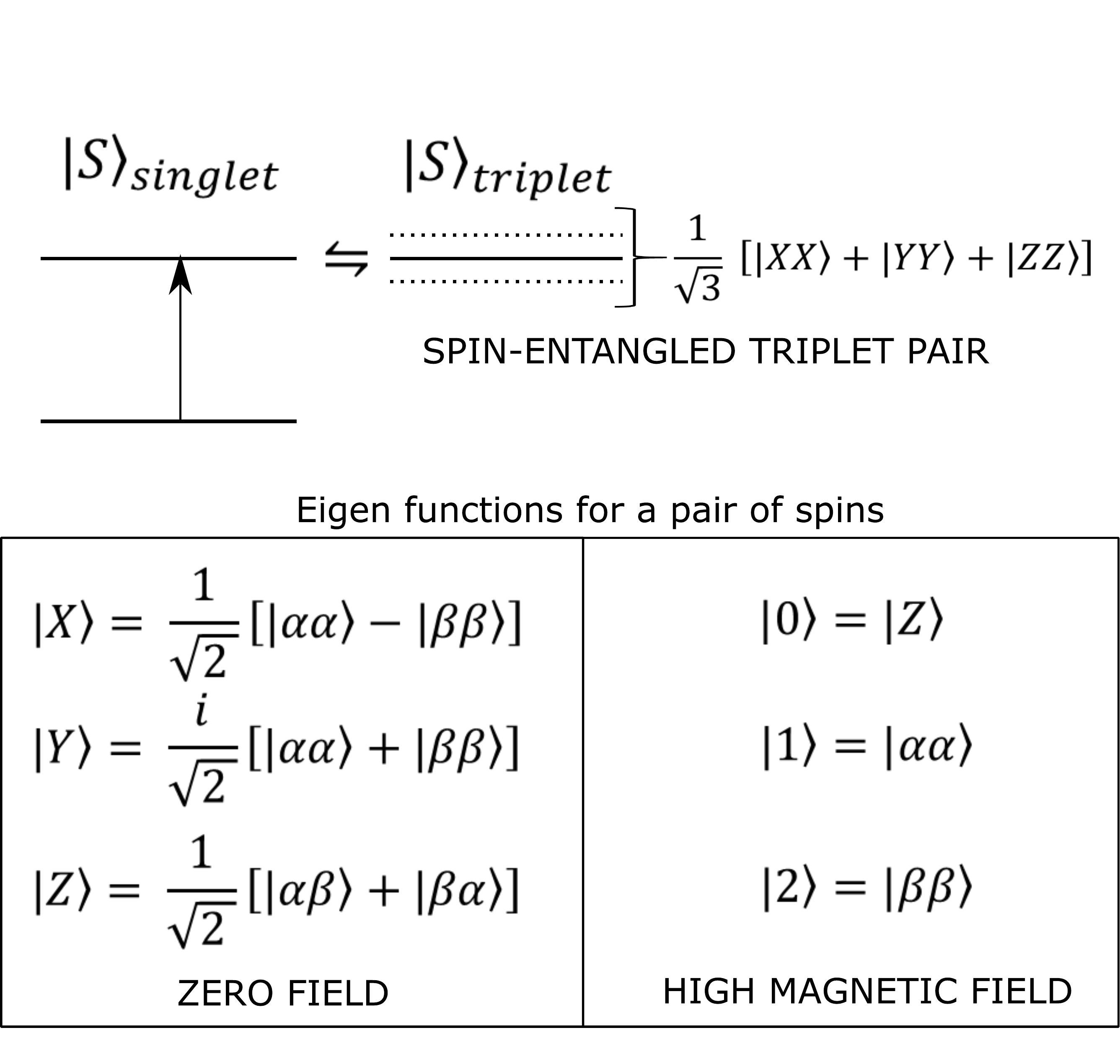}
    \caption{Correlated-triple pair eigen states arising from singlet fission, including magnetic field effects}
    \label{fig:fission}
\end{figure}

Singlet fission and triplet fusion have been investigated extensively over the last few decades with clear consensus that these processes are mediated by a correlated pair of triplets ($|S\rangle_{triplet}$ in Fig.~\ref{fig:fission})~\cite{Smith2010, Chan2011, Grieco2018}. Recently, Yong et al~\cite{Yong2017} proposed that $|S\rangle_{triplet}$ is a spin-entangled quantum state beyond a simple coherent superposition of the free triplet states and can be expanded over spin-basis of individual molecules as shown in Fig.~\ref{fig:fission}. Such a treatment was first put forward by Merrifield~\cite{Merrifield1968}, and later used by Burdett~\emph{et al.}~\cite{Burdett2012} to rationalize coherent oscillations at GHz frequencies observed in time-resolved photoluminescence experiments. Although these earlier works did not specifically address the mediating state as entangled, given that such a representation is indeed the definition of quantum entanglement, Yong \textit{et al}.\  conceptualized their experimental observations accordingly. 
Nevertheless, Merrifield included a cautionary note in his seminal work~\cite{Merrifield1968}: ``\textit{The theory presented here cannot be regarded as confirmed until detailed calculations of field dependence and line shapes have been carried out and compared with experiment.}'' The recent experimental works have only presented indirect albeit strong support for the proposed formalism. Te question remains if a simple superposition of the participating states may be sufficient even without invoking true entanglement. A direct probe of the spin entanglement within $|S\rangle_{triplet}$ is thus both timely as well as crucial in establishing the general mechanism of singlet fission. 

\begin{figure}
    \centering
    \includegraphics[width = 14cm]{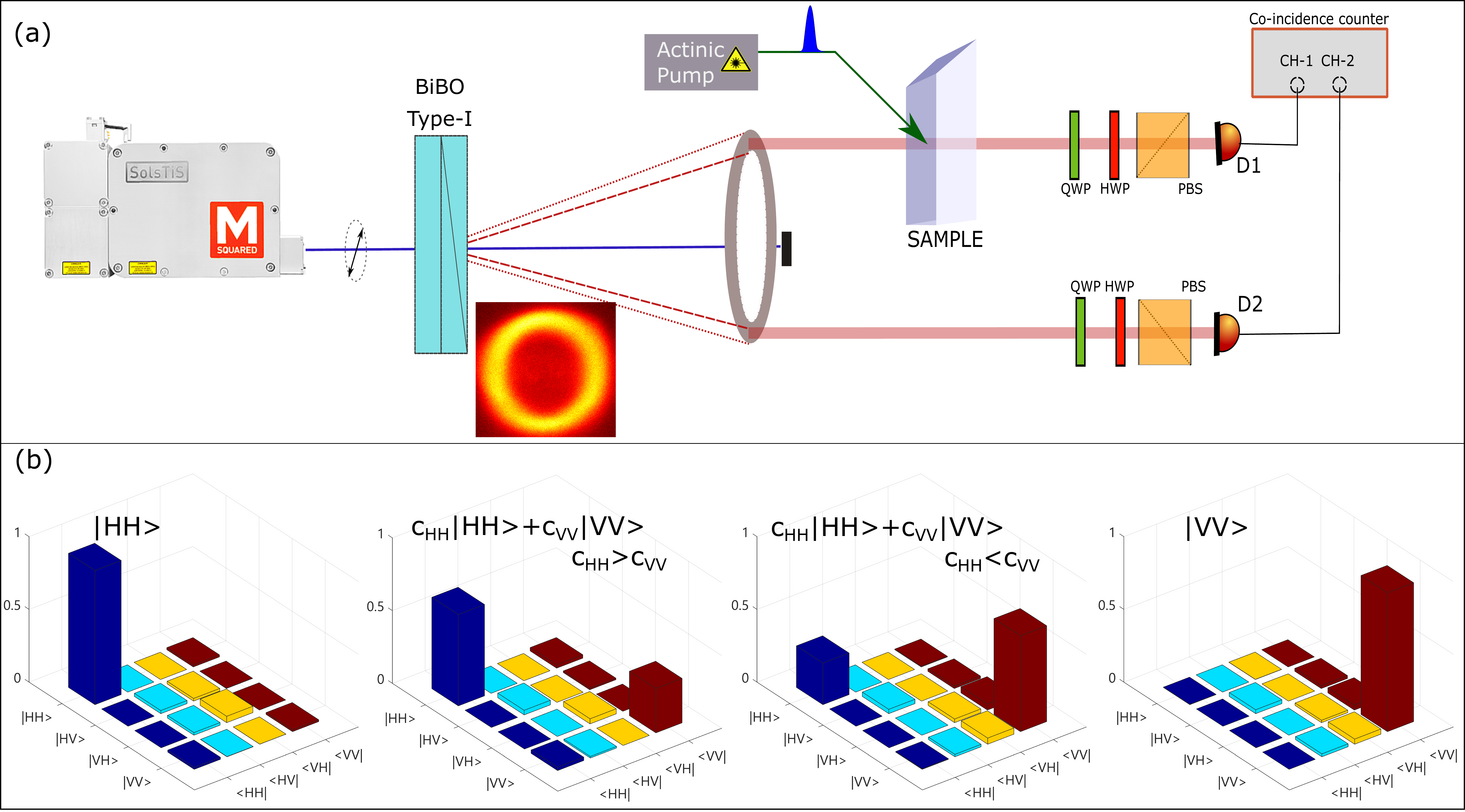}
    \caption{(a) Experimental scheme for quantum process tomography: Narrowband CW laser output at 405\,nm is used to pump a pair of BiBO crystals to generate polarization entangled photons via SPDC (inset shows the image of the SPDC cone of emission). The nature of the entangled state is controlled by tuning the polarization of the pump photons. The signal and idler photons are spatially separated and projected onto various polarization bases using a combination of a quarter waveplate (QWP), half waveplate (HWP) and a linear polarizer (LP). The photons are finally detected using single photon detectors (D1 and D2) and conincidence count rate is obtained via photon-counting electronics. (b) Estimated density matrices for various biphoton states measured via quantum process tomography.}
    \label{fig:spdc}
\end{figure}

Our experimental methodology is based on the quantum tomography apparatus~\cite{Altepeter:2003aa, Altepeter:2005aa} shown in Fig.~\ref{fig:spdc}(a) that  estimates the density matrix of a polarization entangled photon state. The experimental system consists of a photon-generation stage, where the entangled photons at 810\,nm are generated via spontaneous parametric downconversion (SPDC) in a pair of type-I Bismuth Barium Borate (BiBO) crystals~\cite{Rangarajan:2009aa, Kwiat:1999aa}. The pump is a frequency doubled output (at 405\,nm, 50\,mW) of a narrowband Ti:sapphire CW laser (Msquared), whose polarization is controlled by a half-wave plate to obtain different conditions of entangled photons. With any arbitrary pump polarization one obtains a generic entangled state of the form 
$|HH\rangle + e^{i\alpha}|VV\rangle$, where $H$ and $V$ represent horizontal and vertical polarizations of photons respectively. The image of the SPDC emission cone obtained with an EMCCD camera (Andor) is shown as an inset in Fig~\ref{fig:spdc}(a). The signal and idler photons are emitted along two opposite ends of the cone and are spatially separated and sent to two balanced polarimeters. The latter are composed of a quarter waveplate (QWP), half waveplate(HWP) and a polarizing beam splitter(PBS) that project each of the photons onto polarization bases defined by horizontal (H), vertical (V), diagonal (D), anti-diagonal (A),  right (R) or left (L) circular polarizations. 


The photons are then detected via single photon Avalanche photo-diodes and coincidence count rate is recorded via photon counting electronics (Picoquant Hydraharp). The photon-coincidence rates obtained under various polarization projections to construct the density matrix. 

Given that the denisty matrix of a biphoton state is a 4X4 matrix, 16 measurements over the [H,V,D,R] basis is required to estimate it. The formula for tomographic reconstruction of the biphoton density matrix can be written as~\cite{james2005measurement}:
\begin{equation}
    \hat{\rho} = \frac{\sum_{\nu = 1}^{16}\hat{M_{\nu}}n_{\nu}}{\sum_{\nu = 1}^{4}n_{\nu}},
\end{equation}

\begin{figure}
    \centering
    \includegraphics[width=5cm]{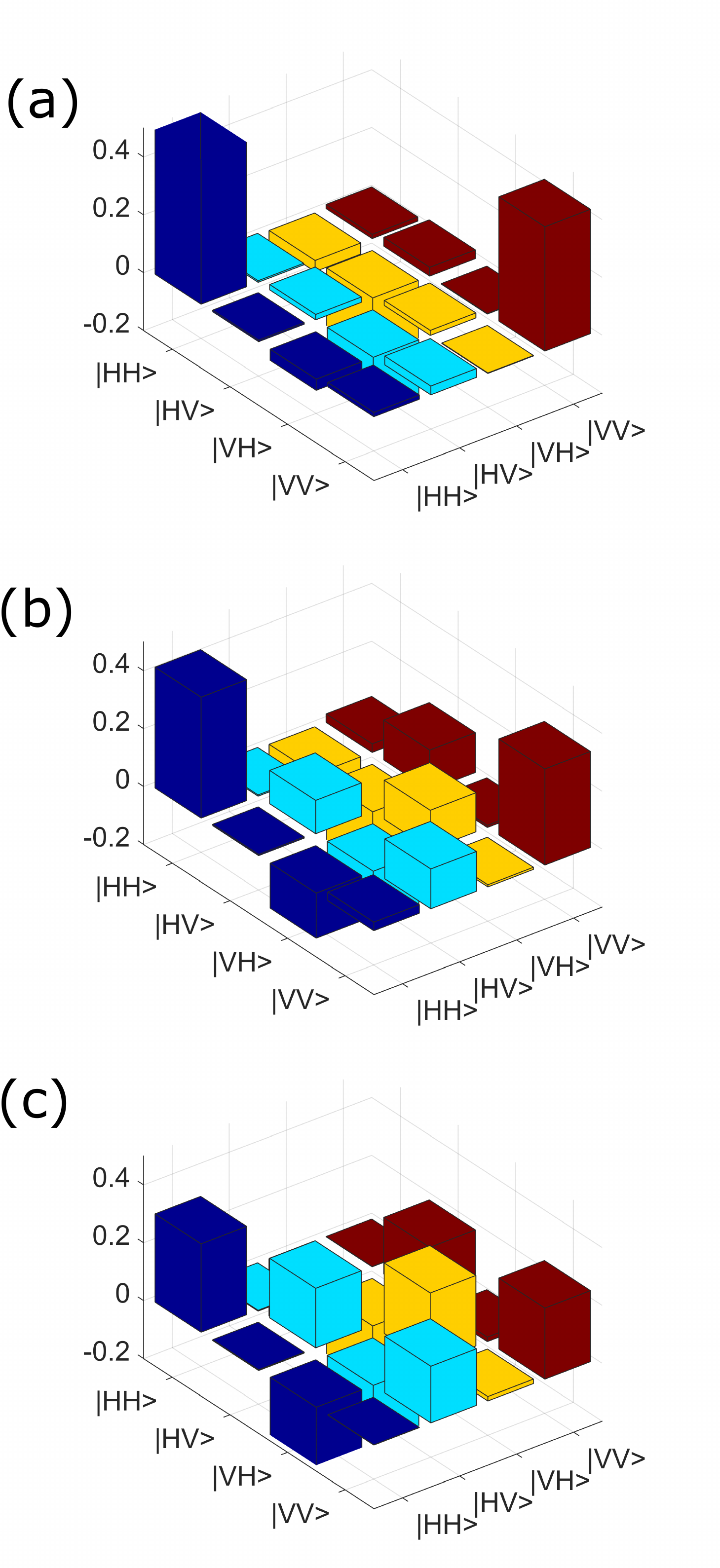}
    \caption{Density matrices with a quarter wave plate placed in the path of one of the photons with the optic axis at (a) 0$^o$, (b) 22.5$^o$ and (c) 45$^o$}
    \label{fig:QWP}
\end{figure}

where $n_{\nu}$ are the experimentally obtained coincidence rates and $\hat{M}_{\nu}$ are a linearly independent set of sixteen 4 X 4 matrices obtained from the two-qubit Pauli matrices ($\hat{\sigma_i}\bigotimes\hat{\sigma_j} (i = 0,1,2,3)$). A few examples of tomographically reconstructed density matrices for various input pump polarizations is shown in Fig.~\ref{fig:spdc}(b). When the input polarization is set at H(V), we obtain biphoton state of the form $|HH\rangle$ ($|VV\rangle$), while an intermediate configuration generates the remaining two. For the sake of our spectroscopic investigation, we set the pump polarization close to 45$^o$ to obtain a maximally entangled state. Note however that at the moment we are unable to generate a pure Bell state. Given our interest in observing the change induced by a sample interaction, this is not detrimental to the experiment. We also note that these matrices are only estimations of the actual density matrix, a minimization algorithm that includes a rigorous error analysis is also being implemented to obtain the maximum likelihood matrix which accurately represents the biphoton state~\cite{james2005measurement}.

Our experimental strategy is to measure the changes induced in the one of the photons via material interactions which will subsequently manifest as a change in the photon density matrix. This can be either due to an unitary transformation of the biphoton state due to a polarization-specific process in the material or a change in the entanglement entropy due to the mixing of the photon state with matter excitations~\cite{cuevas2018first}. In the weak matter-photon coupling limit such as in the systems of interest here, the contribution from the latter is substantially lowered. In such a scenario, the sample interaction may be theoretically captured as a first order scattering amplitude, as ${\cal S}^{(1)}_{\alpha\beta} = \langle \hat\psi_{\alpha,out}^\dagger\hat\psi_{\beta,in} \rangle$ where $\hat \psi_{\beta,in}$ and $\hat \psi_{\alpha,out}^\dagger$ are the Heisenberg operators for removing an incoming photon (with polarization $\beta$) and replacing it with an out-going photon with polarization $\alpha$.  The final output state is thus given as $|out\rangle = {\cal S}^{(1)}$$ |in\rangle$, where $|in\rangle$ is the input entangled state and ${\cal S}^{(1)}$ carries the information about the polarization-selective process in the sample. 

To evaluate the sensitivity of our experimental setup to possible photon transformations, we introduced a quarter-wave plave in the place of the sample (see Fig.~\ref{fig:spdc}(a)) to impose a known retardance in one of the photons. Figs.~\ref{fig:QWP}(a), (b) and (c) show the density matrices when the optical axis of the \textit{QWP-sample} is set to 0$^o$, 22.5$^o$ and 45$^o$ respectively. Clear changes can be observed that are indicative of an unitary transformation, when the Bloch sphere of one of the photons is rotated without the loss of purity in the biphoton state (the traces of the matrices are still approximately close to 1). The changes in the bi-photon density matrix thus contain an imprint of the sample response ${\cal S}^{(1)}$.   

We now return to the problem at hand, which is to probe the nature of the fission intermediate $|S\rangle_{triplet}$ via the entangled photon state. The specific spin-selection rules of the excited state absorption of $|S\rangle_{triplet}$ give us a window into the polarization-specific scattering mechanisms and subsequently to the nature of the quantum interaction. Here we consider a concentrated solution (200 mg/ml) of bis(triisopropylsilylethynyl), (TIPS)-tetracene molecules (obtained via collaboration with Prof. John Anthony, University of Kentucky). TIPS-tetracene at these concentrations has been shown to generate long-living excimer-like states via singlet fission mechanism ~\cite{Stern2015}. We photo-excite the samples with a 80MHz train of picosecond pulses at 500\,nm to generate a steady-state population of these precursors states ($|S\rangle_{triplet}$), which have been proposed to be entangled pairs of triplet excited states ~\cite{Yong2017}. The probe biphoton state in our current experiment at 1.53\,eV is resonant with the excited-state absorption of these states~\cite{Stern2015}. 

\begin{figure}
    \centering
    \includegraphics[width=5cm]{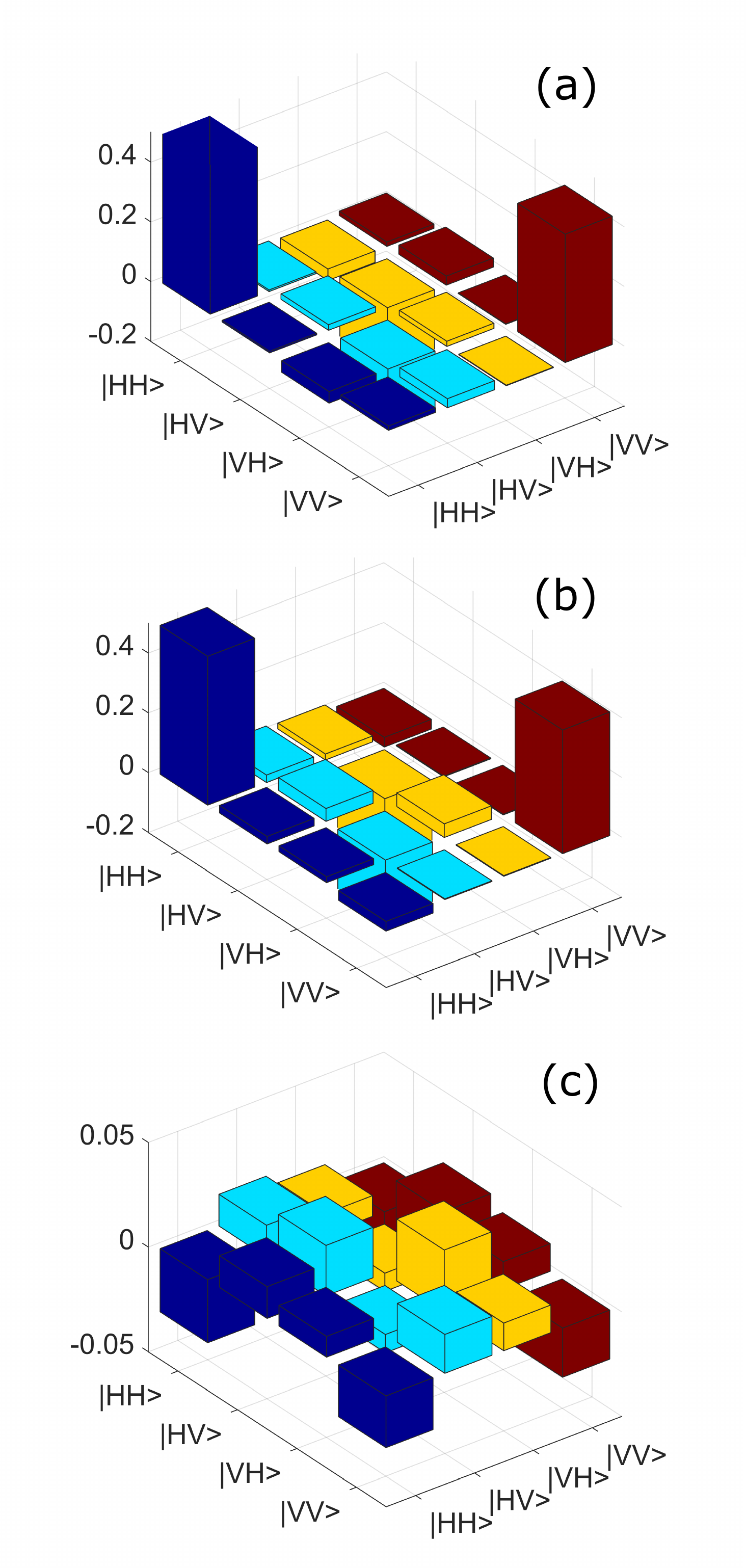}
    \caption{Density matrix obtained (a)without and (b)with a photo-excited sample of concentrated TIPS-tetracene solution in the path of one of the photons. (b) The differential density matrix highlighting the changes induced by the sample in the biphoton state.}
    \label{fig:tetracene}
\end{figure}

Shown in Fig.~\ref{fig:tetracene} (a) and (b) are the density matrices of the biphoton state after transmitting through the sample, but with and without the photo-excited $|S\rangle_{triplet}$ population respectively. Pertinent changes can be observed in the matrix which can be perceived better in the
difference between the matrices (pump on − pump off), plotted in Fig.~\ref{fig:tetracene}(c). Clear non-zero photo-induced components appear in the projections such as $|VH\rangle \langle HV|$ and $|VV\rangle \langle VH|$. Such a transformation is similar to what was observed with the \textit{QWP-sample} shown in Fig.~\ref{fig:QWP}(b) and (c), albeit with substantially lower amplitude. We note that there is a need to perform a systematic analysis considering the fluctuations in the experimental conditions in order to estimate true changes induced by matter interactions. In spite of lack of such an analysis, we consider that the observations presented in Fig.~\ref{fig:tetracene}(c) are indicative of nature of $|S\rangle_{triplet}$ and a comprehensive theoretical treatment will be developed based on it. 

\begin{acknowledgments}
A.R.S.K. acknowledges funding from EU Horizon 2020 via Marie Sklodowska Curie Fellowship (Global) (Project No. 705874).The work in Georgia Tech was supported by the National Science Foundation (Grant No. CHE-1836075). 
\end{acknowledgments}



\begin{thebibliography}{13}%
	\makeatletter
	\providecommand \@ifxundefined [1]{%
		\@ifx{#1\undefined}
	}%
	\providecommand \@ifnum [1]{%
		\ifnum #1\expandafter \@firstoftwo
		\else \expandafter \@secondoftwo
		\fi
	}%
	\providecommand \@ifx [1]{%
		\ifx #1\expandafter \@firstoftwo
		\else \expandafter \@secondoftwo
		\fi
	}%
	\providecommand \natexlab [1]{#1}%
	\providecommand \enquote  [1]{``#1''}%
	\providecommand \bibnamefont  [1]{#1}%
	\providecommand \bibfnamefont [1]{#1}%
	\providecommand \citenamefont [1]{#1}%
	\providecommand \href@noop [0]{\@secondoftwo}%
	\providecommand \href [0]{\begingroup \@sanitize@url \@href}%
	\providecommand \@href[1]{\@@startlink{#1}\@@href}%
	\providecommand \@@href[1]{\endgroup#1\@@endlink}%
	\providecommand \@sanitize@url [0]{\catcode `\\12\catcode `\$12\catcode
		`\&12\catcode `\#12\catcode `\^12\catcode `\_12\catcode `\%12\relax}%
	\providecommand \@@startlink[1]{}%
	\providecommand \@@endlink[0]{}%
	\providecommand \url  [0]{\begingroup\@sanitize@url \@url }%
	\providecommand \@url [1]{\endgroup\@href {#1}{\urlprefix }}%
	\providecommand \urlprefix  [0]{URL }%
	\providecommand \Eprint [0]{\href }%
	\providecommand \doibase [0]{http://dx.doi.org/}%
	\providecommand \selectlanguage [0]{\@gobble}%
	\providecommand \bibinfo  [0]{\@secondoftwo}%
	\providecommand \bibfield  [0]{\@secondoftwo}%
	\providecommand \translation [1]{[#1]}%
	\providecommand \BibitemOpen [0]{}%
	\providecommand \bibitemStop [0]{}%
	\providecommand \bibitemNoStop [0]{.\EOS\space}%
	\providecommand \EOS [0]{\spacefactor3000\relax}%
	\providecommand \BibitemShut  [1]{\csname bibitem#1\endcsname}%
	\let\auto@bib@innerbib\@empty
	\bibitem [{\citenamefont {Smith}\ and\ \citenamefont
		{Michl}(2010)}]{Smith2010}%
	\BibitemOpen
	\bibfield  {author} {\bibinfo {author} {\bibfnamefont {M.~B.}\ \bibnamefont
			{Smith}}\ and\ \bibinfo {author} {\bibfnamefont {J.}~\bibnamefont {Michl}},\
	}\href {\doibase 10.1021/cr1002613} {\bibfield  {journal} {\bibinfo
			{journal} {Chemical reviews}\ }\textbf {\bibinfo {volume} {110}},\ \bibinfo
		{pages} {6891} (\bibinfo {year} {2010})}\BibitemShut {NoStop}%
	\bibitem [{\citenamefont {Chan}\ \emph {et~al.}(2011)\citenamefont {Chan},
		\citenamefont {Ligges}, \citenamefont {Jailaubekov}, \citenamefont {Kaake},
		\citenamefont {Miaja-Avila},\ and\ \citenamefont {Zhu}}]{Chan2011}%
	\BibitemOpen
	\bibfield  {author} {\bibinfo {author} {\bibfnamefont {W.-L.}\ \bibnamefont
			{Chan}}, \bibinfo {author} {\bibfnamefont {M.}~\bibnamefont {Ligges}},
		\bibinfo {author} {\bibfnamefont {A.}~\bibnamefont {Jailaubekov}}, \bibinfo
		{author} {\bibfnamefont {L.}~\bibnamefont {Kaake}}, \bibinfo {author}
		{\bibfnamefont {L.}~\bibnamefont {Miaja-Avila}}, \ and\ \bibinfo {author}
		{\bibfnamefont {X.-Y.}\ \bibnamefont {Zhu}},\ }\href {\doibase
		10.1126/science.1213986} {\bibfield  {journal} {\bibinfo  {journal} {Science
				(New York, N.Y.)}\ }\textbf {\bibinfo {volume} {334}},\ \bibinfo {pages}
		{1541} (\bibinfo {year} {2011})}\BibitemShut {NoStop}%
	\bibitem [{\citenamefont {Grieco}\ \emph {et~al.}(2018)\citenamefont {Grieco},
		\citenamefont {Kennehan}, \citenamefont {Kim}, \citenamefont {Pensack},
		\citenamefont {Brigeman}, \citenamefont {Rimshaw}, \citenamefont {Payne},
		\citenamefont {Anthony}, \citenamefont {Giebink}, \citenamefont {Scholes},\
		and\ \citenamefont {Asbury}}]{Grieco2018}%
	\BibitemOpen
	\bibfield  {author} {\bibinfo {author} {\bibfnamefont {C.}~\bibnamefont
			{Grieco}}, \bibinfo {author} {\bibfnamefont {E.~R.}\ \bibnamefont
			{Kennehan}}, \bibinfo {author} {\bibfnamefont {H.}~\bibnamefont {Kim}},
		\bibinfo {author} {\bibfnamefont {R.~D.}\ \bibnamefont {Pensack}}, \bibinfo
		{author} {\bibfnamefont {A.~N.}\ \bibnamefont {Brigeman}}, \bibinfo {author}
		{\bibfnamefont {A.}~\bibnamefont {Rimshaw}}, \bibinfo {author} {\bibfnamefont
			{M.~M.}\ \bibnamefont {Payne}}, \bibinfo {author} {\bibfnamefont {J.~E.}\
			\bibnamefont {Anthony}}, \bibinfo {author} {\bibfnamefont {N.~C.}\
			\bibnamefont {Giebink}}, \bibinfo {author} {\bibfnamefont {G.~D.}\
			\bibnamefont {Scholes}}, \ and\ \bibinfo {author} {\bibfnamefont {J.~B.}\
			\bibnamefont {Asbury}},\ }\href {\doibase 10.1021/acs.jpcc.7b11228}
	{\bibfield  {journal} {\bibinfo  {journal} {Journal of Physical Chemistry C}\
		}\textbf {\bibinfo {volume} {122}},\ \bibinfo {pages} {2012} (\bibinfo {year}
		{2018})}\BibitemShut {NoStop}%
	\bibitem [{\citenamefont {Yong}\ \emph {et~al.}(2017)\citenamefont {Yong},
		\citenamefont {Musser}, \citenamefont {Bayliss}, \citenamefont {Lukman},
		\citenamefont {Tamura}, \citenamefont {Bubnova}, \citenamefont {Hallani},
		\citenamefont {Meneau}, \citenamefont {Resel}, \citenamefont {Maruyama},
		\citenamefont {Hotta}, \citenamefont {Herz}, \citenamefont {Beljonne},
		\citenamefont {Anthony}, \citenamefont {Clark},\ and\ \citenamefont
		{Sirringhaus}}]{Yong2017}%
	\BibitemOpen
	\bibfield  {author} {\bibinfo {author} {\bibfnamefont {C.~K.}\ \bibnamefont
			{Yong}}, \bibinfo {author} {\bibfnamefont {A.~J.}\ \bibnamefont {Musser}},
		\bibinfo {author} {\bibfnamefont {S.~L.}\ \bibnamefont {Bayliss}}, \bibinfo
		{author} {\bibfnamefont {S.}~\bibnamefont {Lukman}}, \bibinfo {author}
		{\bibfnamefont {H.}~\bibnamefont {Tamura}}, \bibinfo {author} {\bibfnamefont
			{O.}~\bibnamefont {Bubnova}}, \bibinfo {author} {\bibfnamefont {R.~K.}\
			\bibnamefont {Hallani}}, \bibinfo {author} {\bibfnamefont {A.}~\bibnamefont
			{Meneau}}, \bibinfo {author} {\bibfnamefont {R.}~\bibnamefont {Resel}},
		\bibinfo {author} {\bibfnamefont {M.}~\bibnamefont {Maruyama}}, \bibinfo
		{author} {\bibfnamefont {S.}~\bibnamefont {Hotta}}, \bibinfo {author}
		{\bibfnamefont {L.~M.}\ \bibnamefont {Herz}}, \bibinfo {author}
		{\bibfnamefont {D.}~\bibnamefont {Beljonne}}, \bibinfo {author}
		{\bibfnamefont {J.~E.}\ \bibnamefont {Anthony}}, \bibinfo {author}
		{\bibfnamefont {J.}~\bibnamefont {Clark}}, \ and\ \bibinfo {author}
		{\bibfnamefont {H.}~\bibnamefont {Sirringhaus}},\ }\href {\doibase
		10.1038/ncomms15953} {\bibfield  {journal} {\bibinfo  {journal} {Nature
				Communications}\ }\textbf {\bibinfo {volume} {8}},\ \bibinfo {pages} {15953}
		(\bibinfo {year} {2017})}\BibitemShut {NoStop}%
	\bibitem [{\citenamefont {Merrifield}(1968)}]{Merrifield1968}%
	\BibitemOpen
	\bibfield  {author} {\bibinfo {author} {\bibfnamefont {R.~E.}\ \bibnamefont
			{Merrifield}},\ }\href {\doibase 10.1063/1.1669777} {\bibfield  {journal}
		{\bibinfo  {journal} {The Journal of Chemical Physics}\ }\textbf {\bibinfo
			{volume} {48}},\ \bibinfo {pages} {4318} (\bibinfo {year} {1968})},\ \Eprint
	{http://arxiv.org/abs/1011.1669} {arXiv:1011.1669} \BibitemShut {NoStop}%
	\bibitem [{\citenamefont {Burdett}\ and\ \citenamefont
		{Bardeen}(2012)}]{Burdett2012}%
	\BibitemOpen
	\bibfield  {author} {\bibinfo {author} {\bibfnamefont {J.~J.}\ \bibnamefont
			{Burdett}}\ and\ \bibinfo {author} {\bibfnamefont {C.~J.}\ \bibnamefont
			{Bardeen}},\ }\href {\doibase 10.1021/ja301683w} {\bibfield  {journal}
		{\bibinfo  {journal} {Journal of the American Chemical Society}\ }\textbf
		{\bibinfo {volume} {134}},\ \bibinfo {pages} {8597} (\bibinfo {year}
		{2012})}\BibitemShut {NoStop}%
	\bibitem [{\citenamefont {Altepeter}\ \emph {et~al.}(2003)\citenamefont
		{Altepeter}, \citenamefont {Branning}, \citenamefont {Jeffrey}, \citenamefont
		{Wei}, \citenamefont {Kwiat}, \citenamefont {Thew}, \citenamefont {O'Brien},
		\citenamefont {Nielsen},\ and\ \citenamefont {White}}]{Altepeter:2003aa}%
	\BibitemOpen
	\bibfield  {author} {\bibinfo {author} {\bibfnamefont {J.~B.}\ \bibnamefont
			{Altepeter}}, \bibinfo {author} {\bibfnamefont {D.}~\bibnamefont {Branning}},
		\bibinfo {author} {\bibfnamefont {E.}~\bibnamefont {Jeffrey}}, \bibinfo
		{author} {\bibfnamefont {T.~C.}\ \bibnamefont {Wei}}, \bibinfo {author}
		{\bibfnamefont {P.~G.}\ \bibnamefont {Kwiat}}, \bibinfo {author}
		{\bibfnamefont {R.~T.}\ \bibnamefont {Thew}}, \bibinfo {author}
		{\bibfnamefont {J.~L.}\ \bibnamefont {O'Brien}}, \bibinfo {author}
		{\bibfnamefont {M.~A.}\ \bibnamefont {Nielsen}}, \ and\ \bibinfo {author}
		{\bibfnamefont {A.~G.}\ \bibnamefont {White}},\ }\href {\doibase
		10.1103/PhysRevLett.90.193601} {\bibfield  {journal} {\bibinfo  {journal}
			{Physical Review Letters}\ }\textbf {\bibinfo {volume} {90}},\ \bibinfo
		{pages} {4} (\bibinfo {year} {2003})},\ \Eprint
	{http://arxiv.org/abs/0303038} {arXiv:0303038 [quant-ph]} \BibitemShut
	{NoStop}%
	\bibitem [{\citenamefont {Altepeter}, \citenamefont {Jeffrey},\ and\
		\citenamefont {Kwiat}(2005)}]{Altepeter:2005aa}%
	\BibitemOpen
	\bibfield  {author} {\bibinfo {author} {\bibfnamefont {J.~B.}\ \bibnamefont
			{Altepeter}}, \bibinfo {author} {\bibfnamefont {E.~R.}\ \bibnamefont
			{Jeffrey}}, \ and\ \bibinfo {author} {\bibfnamefont {P.~G.}\ \bibnamefont
			{Kwiat}},\ }\href {\doibase 10.1016/S1049-250X(05)52003-2} {\bibfield
		{journal} {\bibinfo  {journal} {Advances in Atomic, Molecular and Optical
				Physics}\ }\textbf {\bibinfo {volume} {52}},\ \bibinfo {pages} {105}
		(\bibinfo {year} {2005})}\BibitemShut {NoStop}%
	\bibitem [{\citenamefont {Rangarajan}, \citenamefont {Goggin},\ and\
		\citenamefont {Kwiat}(2009)}]{Rangarajan:2009aa}%
	\BibitemOpen
	\bibfield  {author} {\bibinfo {author} {\bibfnamefont {R.}~\bibnamefont
			{Rangarajan}}, \bibinfo {author} {\bibfnamefont {M.}~\bibnamefont {Goggin}},
		\ and\ \bibinfo {author} {\bibfnamefont {P.}~\bibnamefont {Kwiat}},\ }\href
	{\doibase 10.1364/OE.17.018920} {\bibfield  {journal} {\bibinfo  {journal}
			{Optics Express}\ }\textbf {\bibinfo {volume} {17}},\ \bibinfo {pages}
		{18920} (\bibinfo {year} {2009})}\BibitemShut {NoStop}%
	\bibitem [{\citenamefont {Kwiat}\ \emph {et~al.}(1999)\citenamefont {Kwiat},
		\citenamefont {Waks}, \citenamefont {White}, \citenamefont {Appelbaum},\ and\
		\citenamefont {Eberhard}}]{Kwiat:1999aa}%
	\BibitemOpen
	\bibfield  {author} {\bibinfo {author} {\bibfnamefont {P.~G.}\ \bibnamefont
			{Kwiat}}, \bibinfo {author} {\bibfnamefont {E.}~\bibnamefont {Waks}},
		\bibinfo {author} {\bibfnamefont {A.~G.}\ \bibnamefont {White}}, \bibinfo
		{author} {\bibfnamefont {I.}~\bibnamefont {Appelbaum}}, \ and\ \bibinfo
		{author} {\bibfnamefont {P.~H.}\ \bibnamefont {Eberhard}},\ }\href {\doibase
		10.1103/PhysRevA.60.R773} {\bibfield  {journal} {\bibinfo  {journal}
			{Physical Review A}\ }\textbf {\bibinfo {volume} {60}},\ \bibinfo {pages}
		{R773} (\bibinfo {year} {1999})}\BibitemShut {NoStop}%
	\bibitem [{\citenamefont {James}\ \emph {et~al.}(2005)\citenamefont {James},
		\citenamefont {Kwiat}, \citenamefont {Munro},\ and\ \citenamefont
		{White}}]{james2005measurement}%
	\BibitemOpen
	\bibfield  {author} {\bibinfo {author} {\bibfnamefont {D.~F.}\ \bibnamefont
			{James}}, \bibinfo {author} {\bibfnamefont {P.~G.}\ \bibnamefont {Kwiat}},
		\bibinfo {author} {\bibfnamefont {W.~J.}\ \bibnamefont {Munro}}, \ and\
		\bibinfo {author} {\bibfnamefont {A.~G.}\ \bibnamefont {White}},\ }in\
	\href@noop {} {\emph {\bibinfo {booktitle} {Asymptotic Theory of Quantum
				Statistical Inference: Selected Papers}}}\ (\bibinfo  {publisher} {World
		Scientific},\ \bibinfo {year} {2005})\ pp.\ \bibinfo {pages}
	{509--538}\BibitemShut {NoStop}%
	\bibitem [{\citenamefont {Cuevas}\ \emph {et~al.}(2018)\citenamefont {Cuevas},
		\citenamefont {Carre{\~n}o}, \citenamefont {Silva}, \citenamefont
		{De~Giorgi}, \citenamefont {Su{\'a}rez-Forero}, \citenamefont {Mu{\~n}oz},
		\citenamefont {Fieramosca}, \citenamefont {Cardano}, \citenamefont
		{Marrucci}, \citenamefont {Tasco} \emph {et~al.}}]{cuevas2018first}%
	\BibitemOpen
	\bibfield  {author} {\bibinfo {author} {\bibfnamefont {{\'A}.}~\bibnamefont
			{Cuevas}}, \bibinfo {author} {\bibfnamefont {J.~C.~L.}\ \bibnamefont
			{Carre{\~n}o}}, \bibinfo {author} {\bibfnamefont {B.}~\bibnamefont {Silva}},
		\bibinfo {author} {\bibfnamefont {M.}~\bibnamefont {De~Giorgi}}, \bibinfo
		{author} {\bibfnamefont {D.~G.}\ \bibnamefont {Su{\'a}rez-Forero}}, \bibinfo
		{author} {\bibfnamefont {C.~S.}\ \bibnamefont {Mu{\~n}oz}}, \bibinfo {author}
		{\bibfnamefont {A.}~\bibnamefont {Fieramosca}}, \bibinfo {author}
		{\bibfnamefont {F.}~\bibnamefont {Cardano}}, \bibinfo {author} {\bibfnamefont
			{L.}~\bibnamefont {Marrucci}}, \bibinfo {author} {\bibfnamefont
			{V.}~\bibnamefont {Tasco}},  \emph {et~al.},\ }\href@noop {} {\bibfield
		{journal} {\bibinfo  {journal} {Science advances}\ }\textbf {\bibinfo
			{volume} {4}},\ \bibinfo {pages} {eaao6814} (\bibinfo {year}
		{2018})}\BibitemShut {NoStop}%
	\bibitem [{\citenamefont {Stern}\ \emph {et~al.}(2015)\citenamefont {Stern},
		\citenamefont {Musser}, \citenamefont {Gelinas}, \citenamefont {Parkinson},
		\citenamefont {Herz}, \citenamefont {Bruzek}, \citenamefont {Anthony},
		\citenamefont {Friend},\ and\ \citenamefont {Walker}}]{Stern2015}%
	\BibitemOpen
	\bibfield  {author} {\bibinfo {author} {\bibfnamefont {H.~L.}\ \bibnamefont
			{Stern}}, \bibinfo {author} {\bibfnamefont {A.~J.}\ \bibnamefont {Musser}},
		\bibinfo {author} {\bibfnamefont {S.}~\bibnamefont {Gelinas}}, \bibinfo
		{author} {\bibfnamefont {P.}~\bibnamefont {Parkinson}}, \bibinfo {author}
		{\bibfnamefont {L.~M.}\ \bibnamefont {Herz}}, \bibinfo {author}
		{\bibfnamefont {M.~J.}\ \bibnamefont {Bruzek}}, \bibinfo {author}
		{\bibfnamefont {J.}~\bibnamefont {Anthony}}, \bibinfo {author} {\bibfnamefont
			{R.~H.}\ \bibnamefont {Friend}}, \ and\ \bibinfo {author} {\bibfnamefont
			{B.~J.}\ \bibnamefont {Walker}},\ }\href {\doibase 10.1073/pnas.1503471112}
	{\bibfield  {journal} {\bibinfo  {journal} {Proceedings of the National
				Academy of Sciences}\ }\textbf {\bibinfo {volume} {112}},\ \bibinfo {pages}
		{7656} (\bibinfo {year} {2015})}\BibitemShut {NoStop}%
\end{thebibliography}
%

\end{document}